# Stock Price Prediction Using Time Series, Econometric, Machine Learning, and Deep Learning Models

Ananda Chatterjee
*Department of Data Science*
*Praxis Business School*
Kolkata, India
ananda.chatterjee89@gmail.com

Hrisav Bhowmick
*Department of Data Science*
*Praxis Business School*
Kolkata, India
hrisavbhowmick@gmail.com

Jaydip Sen
*Department of Data Science*
*Praxis Business School*
Kolkata, India
jaydip.sen@acm.org

*Abstract*—For a long-time, researchers have been developing a reliable and accurate predictive model for stock price prediction. According to the literature, if predictive models are correctly designed and refined, they can painstakingly and faithfully estimate future stock values. This paper demonstrates a set of time series, econometric, and various learning-based models for stock price prediction. The data of Infosys, ICICI, and SUN PHARMA from the period of January 2004 to December 2019 was used here for training and testing the models to know which model performs best in which sector. One time series model (Holt-Winters Exponential Smoothing), one econometric model (ARIMA), two machine Learning models (Random Forest and MARS), and two deep learning-based models (simple RNN and LSTM) have been included in this paper. MARS has been proved to be the best performing machine learning model, while LSTM has proved to be the best performing deep learning model. But overall, for all three sectors - IT (on Infosys data), Banking (on ICICI data), and Health (on SUN PHARMA data), MARS has proved to be the best performing model in sales forecasting.

I. INTRODUCTION

For a long time, future Stock Price prediction has piqued the interest of researchers. While adherents of the efficient market hypothesis assert that precisely forecasting stock prices is impossible, many studies disagree. There are claims in the literature that show that predictive models, when correctly developed and optimized, may predict future stock prices very precisely and reliably. It's also been discovered that the selection of variables used to build a predictive model, the methods used, and how the model was optimized all affect its accuracy. Researchers proposed a method for predicting stock prices based on the time series decomposition in this regard [1-4]. Machine learning and deep learning have also become increasingly prominent while predicting stock prices and tracking their pattern movements [5-7]. A Time-Series decomposition-based approach was also used to conduct predictive analysis of the FMCG, realty, large-cap, mid-cap, banking, and other sectors of India [8-10]. Researchers have also developed a robust and well-founded predictive paradigm for stock price forecasting of Google using RNN based deep learning technique [11]. Some works have also demonstrated efficacy and a high level of precision in forecasted stock prices using *convolutional neural network (CNN)* and *long-and-short-term memory (LSTM)* network-based models [12-19].

An interaction effect between the *exchange rate* and *stock return* was analyzed by the *quantile regression* approach [20]. Stock price prediction based on the error model and Granger causality test was also introduced by researchers [21]. Stock price prediction on event-based trading, using neural language processing on the news items on the social web, and applying machine learning and deep learning models have also been proposed in the literature [22-23].

The present study encompasses a set of time series (TS), econometric, and learning-based models to predict the future prices of three important stocks of the National Stock Exchange (NSE) of India. The stocks studied by us are Infosys, ICICI, and Sun Pharma. The study is conducted on three sectors: IT, Banking, and Health. For conducting the research of the IT sector, stock price data of Infosys has been taken; for the banking sector, ICICI's stock price data has been chosen; and for the health sector, stock price data of SUN PHARMA has been considered. All three of the aforementioned industries' data are collected between January 2004 and December 2019. All the used models are trained on the data from January 2004 to December 2018, and the performance of the models is tested on the dataset from January 2019 to December 2019. The response variable chosen is *close*. In this study, six models are built, which are Holt-Winter Exponential Smoothing, ARIMA, Random Forest, MARS, RNN, and LSTM. The predictive abilities of the schemes are augmented by introducing a deep learning LSTM model. All the models have different architecture, and their theory of operation is different from one another. The time series, econometric, and deep learning-based models had univariate input data, while the machine learning models had multivariate input data.

The major contributions of our works are as follows. First, we have developed six models combining time series, econometric, and learning-based techniques and applied those to three major sectors for stock price prediction with very high precision. Second, using the LSTM model, we forecast the stock price of the eighth day based on the past seven days' stock values, and finally, we have been able to figure out among the applied models which model works best in which sector. The best-performing model is identified base on the lowest value of the ratio of the RMSE and the mean value of *close* of stock price.

The paper has been organized into five sections. The problem statement that we intend to solve in this paper has been defined in Section II. Section III gives a quick review of some related stock price forecasting research. The methodology that has been adopted in solving the problem has been illustrated in Section IV. The experimental results of different approaches taken have been shown in Section V, and in Section VI, the work has been wrapped up, and some potential future research directions have been provided.

## II. PROBLEM STATEMENT

The goal of our research is to build a robust model to predict stock prices of three different sectors IT, Banking, and Health. For this purpose, over 15 years of data from Jan 2004-Dec, 2019 have been acquired from Yahoo Finance for three organizations such as Infosys, ICICI, and SUN PHARMA, which fall under the IT, Banking, and Health sector, respectively. A gamut of time series, econometric, machine learning, and deep learning models were implemented to forecast stock values of the above-mentioned three organizations. We hypothesize that, among the employed ML models, the MARS model will provide the best accuracy because of its capacity to identify significant features from a dataset and develop a model that is a combination of a set of linear functions. It is also believed that the LSTM model will produce outstanding results due to its capacity to extract a huge number of features from data.

## III. RELATED WORK

The current state of stock price prediction research in the paper can be divided into three groups. In the first part of the work, the technique Holt-Winters Exponential Smoothing deals with univariate data, which works well in producing short time forecasts, but it has shortcomings, including the normalization of seasonal indices, the choice of the starting values, and the choice of smoothing parameters [24]. The second category of the work uses the technique of ARIMA, but the only disadvantage associated with the method is the series needs to be converted to stationary before applying ARIMA. Moreover, the assumption of the ARIMA model is constant variance, but financial time series exhibit changes in volatility, and this feature doesn't come under the ARIMA assumption [25]. The models which fall under the third category involve machine learning techniques such as Random Forest and MARS. Random Forest regression sometimes falls short in predicting while there is randomness in data, but MARS improves the regression problem output by automatic feature selection, i.e., removing those predictor variables which do not contribute to the model [26-27]. The fourth category in our study includes deep learning-based models such as RNN and LSTM, which can learn the nonlinear pattern from the past data; so, randomness in financial time series can easily be handled, resulting in obtaining a highly precise forecasted stock value [28-29].

## IV. METHODOLOGY

In this work, the methods which we follow to predict stock prices are time series, econometric, and learning-based approaches. The time series includes one model Holt-Winter Exponential Smoothing, the econometric method involves one model, ARIMA, and the machine learning technique involves two models Random Forest and MARS, and the deep learning method involves two models RNN and LSTM. All the above-stated models were applied to three important sectors for stock price forecasting such as IT, Banking, and Health. The stock value of three firms, including Infosys, ICICI, and SUN PHARMA, has been used to test these models. For the period of January 2004 to December 2019, stock prices for all three firms were obtained from the Yahoo Finance website. The Stock values consist of the following variables: (a) *date*, (b) *open*, (c) *high*, (d) *low*, and (e) *close*. For all the models, the *close* value of stock price has been taken as the target variable. The working principle of each model has been discussed below.

### A. Holt-Winters Exponential Smoothing

This method takes into account the weighted averages of prior data, where the weights are decreasing exponentially as the observations grow older. Single exponential smoothing captures single parameter *alpha*, which captures the level of the time series. Double Exponential smoothing, which is also called the Holts Exponential smoothing, captures level and trend. In Holts Winter Exponential Smoothing, along with level and trend, seasonality in the data is also captured. Here a parameter called *gamma* is added, along with *alpha* and *beta*, to control the influence of the seasonal component. The capturing of the three parameters results in a higher accuracy value in forecasting.

In our case, a univariate analysis was done using Holt-Winters Exponential Smoothing, applying it on the *close* column of stock prices of three companies. Initially, there were 3964 records, and after removing the null values, it turned into 3949 records. The data of INFOSYS, ICICI, and SUN PHARMA from January 2004 to December 2018 (3708 records) was used as a training dataset, and the data of 2019 (241 records) was used as a testing dataset. An *Exponential Smoothing* () function was imported from the *statsmodels package*. The *seasonal* parameter was set as 'additive' as seasonality was constant in the dataset. The *seasonal periods* were chosen as 5, which means the data will exhibit seasonality after every five days. The ratio of RMSE score and mean of test part of the *close* was computed to determine how well the model performs. Among three sectors Holt-Winters Exponential Smoothing model performed best in the case of the health sector (on SUN PHARMA data) and performed poorly for the banking sector (on ICICI data).

### B. ARIMA

Auto-Regressive Integrated Moving Average (ARIMA) comprises of three terms *autoregression* (AR) which means the forecasted value depends on linear combinations of own past observed values. *Integrated* (I) means differencing of the time series to make it stationary. In a non-stationary time series, due to the effect of trend and seasonality, the prediction may be inaccurate. So, to avoid this, the time series is made stationary first by differencing, which is nothing but the difference between the observations and recent past observations. The stationarity behavior of the series is studied using the Augmented Dickey-Fuller (ADF) and Kwiatkowski–Phillips–Schmidt–Shin (KPSS) test [30]. *Moving average* (MA), which takes the average of residuals of the past few observations to forecast the present value. The moving average term nullifies if any error due to noise incorporates in the data. The ARIMA consists of three parameters $p$, $d$, and $q$, where $p$ refers to the maximum lag to consider in forecasting, $d$ is the differencing needed to achieve a stationary series, and $q$ refers to the maximum number of errors to be considered in prediction.

In our case, a univariate analysis was done using ARIMA, applying it on the *close* column of stock prices of three companies. Initially, there were 3964 records, and after removing the null values, it turned into 3949 records. The data of Infosys, ICICI, and SUN PHARMA from January 2004 to December 2018 (3708 records) was used as a training dataset, and the data of 2019 (241 records) was used as a testing dataset. The *ARIMA()* function was imported from the *arima_model* sub-package of *statsmodel.tsa* package. The auto_arima function was run to determine the optimum values of p, d, and q. The values of *p*, *d*, and *q* were obtained as 3, 1, 1 with the lowest value of AIC = 24234.323 for data of INFOSYS. The values of p, d, and q were obtained as 2, 1, 2 with the lowest value of AIC = 23119.75 for data of ICICI. The values of p, d, and q were obtained as 2, 1, 1 with the lowest value of AIC = 26685.265 for data of SUN PHARMA. After obtaining the optimum value of p, d, and q for a particular sector, it was fed as an input to the *ARIMA()* function to get the forecasted stock price value. Ratio RMSE score and test mean of the *close* was calculated when comparing y-test and y-pred to determine the model's performance. Among three sectors, ARIMA performed well for both the IT (Infosys) and banking (ICICI) sector and performed poorly for the healthcare sector (SUN PHARMA).

*C. RANDOM FOREST*

Random Forest Regression is an ensemble technique based on a collection of decision trees to predict output for each of the trees and then finally takes the average of all the predictions, which is regarded as the output prediction of the random forest model. It uses the concept of bagging, i.e., Bootstrap and Aggregation. Bootstrap means choosing random samples with replacement from the dataset. Aggregation is the combination of all the predictions to get the final output. Bagging helps to reduce the overfitting of the models.

For model building purposes, a multivariate analysis was done on the variables *open*, *high*, *low*, *close*, *adj-close*, and *volume*. Infosys, ICICI, SUN PHARMA data from 2004 to 2018 (3708 records) was used as a training dataset, while data of 2019 (241 records) was selected for the test dataset. *MinMax* scaling was applied to bring all the features to a range between 0 and 1. The random forest model was fitted on *X-train* and *y-train*, and the values of the *close* column were predicted on the basis of the *X-test*. Values were inversely scaled to the original scale. The ratio of RMSE score and test of *close* mean as well as the train of *close* mean was calculated when comparing *y-test* and *y-pred*. Among the three industries, the model performed best on the health sector (SUN PHARMA data) and performed poorly on the Banking Sector (ICICI data) based on RMSE/Test Mean ratio.

*D. MARS*

Multivariate Adaptive Regression Spline (MARS) adapts the nonlinearity from the dataset. The MARS algorithm works by splitting the input variables into several step functions. These are known as basis functions. These functions are assessed from the cut points of the data known as *knots*. At each knot, the algorithm searches for a range of values and selects the step function that generates the least error values. Then, at each of the knots, a hinge function is applied, and the technique is repeated to create a resilient nonlinear prediction model. The training datasets will fit better if the number of knots is increased, but it might lead to overfitting. An overfitted model will cause less reliable test data results, necessitating model pruning using cross-validation to reach the ideal number of knots. MARS' concept of operation entails creating a complex model by combining step functions in pairs at each knot. During the forward pass phase, the knots must be identified. The backward pass phase is when the algorithm tries to trim or delete the terms that are poor contributors in order to avoid overfitting.

For model building purposes, a multivariate analysis was done on *open*, *high*, *low*, *close*, *adj-close*, and *volume*. The data of Infosys, ICICI, SUN PHARMA from 2004 to 2018 (3708 records) was used as a training dataset, while data of 2019 (241 records) was selected for the test dataset. *Minmax* scaling was applied to bring all the features to a range between 0 and 1. The *earth*() function was imported from the *py-earth package* to run this MARS model. The MARS model was fitted on *X-train* and *y-train*, and the values of the *close* column were predicted on the basis of the *X-test*. Values were inversely scaled to the original scale. The ratio of RMSE score and test means of the *close* was calculated when comparing y-test and y-pred. Among the three industries, the model performed best in the case of the banking Sector (on ICICI data) and performed poorly in the case of the health Sector (on SUN PHARMA data).

*E. RNN and LSTM*

For deep learning, two models – Simple RNN and Stacked LSTM – were used for model building. The data of Infosys, ICICI, SUN PHARMA from 2004 to 2018 (3708 records) was used as a training dataset, while data of 2019 (241 records) was selected for the test dataset. A univariate analysis was done, taking only the *close* column in scope. First, it was scaled using *MinMaxScaler* from a range of 0 to 1. Then the training and testing data was prepared. For this, seven days rolling window was used as a time step. The objective was to make *X-train* on the first seven days (index: 0 to 6) of the train set. *X-train* will be an array of scaled values. Then on the basis of that, the 8th day's (index: 7) value will be *y-train*. Then the sliding window will move forward by one day. Again, *X-train* will have array values for the next seven days (index: 1 to 7), and the y-train will have the 9th day's (index: 8) value. Next, *X-train* will have array values for index: 2 to 8, and the y-train will have index: 9's value. Like this, it will continue till the *X-train* will have values till the second last record of the train set, and the *y-train* will have the last index of the train set. Finally, the train set will have 3700 days. Next, the counter will be shifted to the testing set. Again, on the test set first seven day's value will be assigned as *X-test,* and the 8th day's value will be the *y-test*. This will go on till the test set gets over. Finally, the test set will have 233 days.

Simple models of machine learning models execute on the principle of sending data in a one-by-one sequence, and prediction is made only on the basis of that. But in the case of RNN or LSTM, we predict values based on past records as well. Here the sequencing is important. RNN works on the elements stored in the short-term memory. These elements help in the sequence prediction. It has at least one feedback connection, which helps in looping activations. For model building, in the first case, we go for the simple RNN model. It will have a sequential of RNN layers followed by a dropout layer. Finally, a *dense layer* is used, which is a deeply connected neural network layer. The input layer of RNN has a number of time steps as seven and a number of features as 1 (which is the *close* column). First hidden RNN layer (Simple-RNN) has 256 nodes, followed by a 20% dropout (dropout). The second hidden RNN layer (simple-rnn-1) has 128 nodes, again followed by a 20% dropout (dropout-1). Tanh activation function was used in both the RNN layers. And finally, a dense output layer (dense). The model is trained on the training set

(X-train and y-train). The model is compiled with the help of Adam optimizer, and the error is computed using Root Mean Squared Error (RMSE). The network is trained for 100 epochs with a batch size of 64. Each of the close values is predicated on the previous seven days of trained data. Among the three industries, the RNN model performed best on the SUN PHARMA data and performed poorly on the ICICI data based on RMSE/Test Mean ratio.

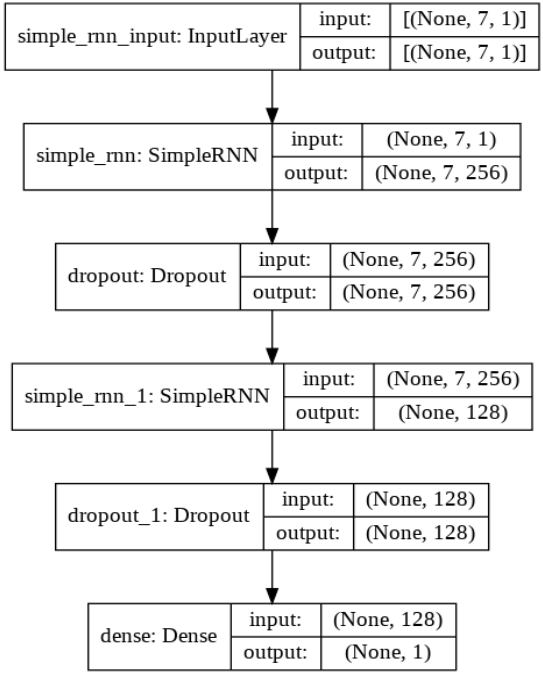

Fig. 1. The RNN model for stock price prediction

However, simple RNN has some shortcomings. It cannot remember memories that are too old. Also, it faces a *vanishing gradient* problem, i.e., the gradient almost has no effect during backpropagation due to the presence of so many layers. LSTM handles this issue well. LSTMs are capable of learning long-term dependencies. It contains a simple RNN cell, cell state (used for long-term memory), and three gates – forget, input and output. Forget gate basically determines which data to keep and which one to remove from memory. Input gate decides till what information is required for the Internal Cell State. Output gate decides on which output to keep from Internal Cell State. In our LSTM model, the input layer is the same as RNN, i.e., number of time steps as seven and number of features as 1 (which is the *close* column). First hidden LSTM layer (*lstm-2*) has 256 nodes, followed by a 20% dropout (*dropout-2*). The second hidden LSTM layer (*lstm-3*) has 128 nodes, again followed by a 20% dropout (*dropout-3*). *Tanh* activation function was used in both the LSTM layers. And finally, a dense output layer (*dense*). The model is compiled with the help of Adam optimizer, and the error is computed using RMSE. The network is trained for 100 epochs with a batch size of 64. Each of the *close* values is predicated on the previous seven days of trained data. Among the three industries, the LSTM model also performed best on the Infosys data and performed poorly on the ICICI data based on RMSE/Test Mean ratio.

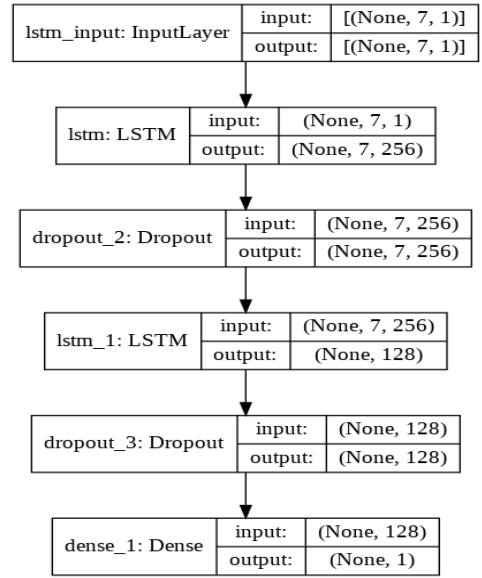

Fig. 2. The LSTM model for stock price prediction

The train set had 3708 records (2004 to 2018), while the test set had 241 records (2019). Using seven days rolling window, we truncated the train set to 3700 days, and the test set was truncated to 233 days. The prediction was made on the truncated values using Simple RNN and Stacked LSTM.

## V. EXPERIMENTAL RESULTS

The performance result demonstrates the efficacy of six models built to forecast the stock price of three different sectors like IT, Bank, and Health. The performance of the models was measured using an evaluation metric that determines the ratio of RMSE and test mean of *close* values of the stock price as well as the train mean of *close* values. For the train set, the mean was the average of *close* values of train data. For the test set, the mean was the average of *close* values of test data. Below are the performance results of each model used in each sector.

### A. IT Sector

In the IT sector, the *close* value of the stock price of Infosys was taken. Six models were applied to get the forecasted stock price. The output of all the models is shown along with the table of each model, comprising the evaluation metric.

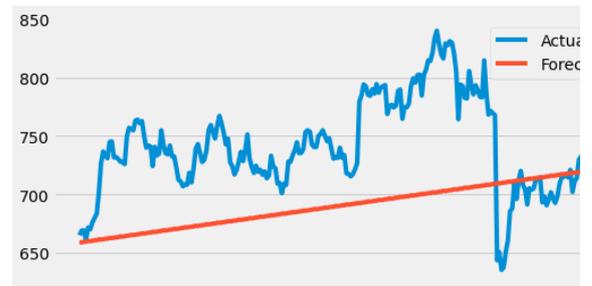

Fig. 3 (a) Actual and Forecasted Stock Prices of Infosys using Holt-Winters Exponential Smoothing

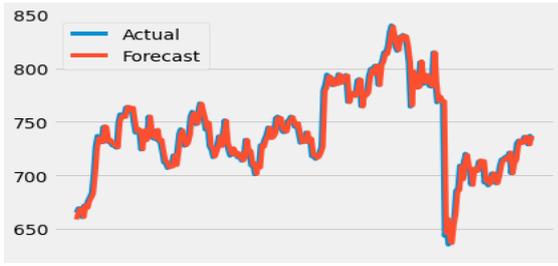

Fig. 3 (b) Actual and Forecasted Stock Prices of Infosys using ARIMA

Table III. Results of Actual and Forecasted Stock Prices of Infosys using TS and econometric based models

| Models | Training Data | | Testing Data | |
|---|---|---|---|---|
| Holt-Winters | RMSE/ mean of *close* | 0.1018 | RMSE/ mean of *close* | 0.089 |
| ARIMA | RMSE/ mean of *close* | 0.033 | RMSE/ mean of *close* | 0.018 |

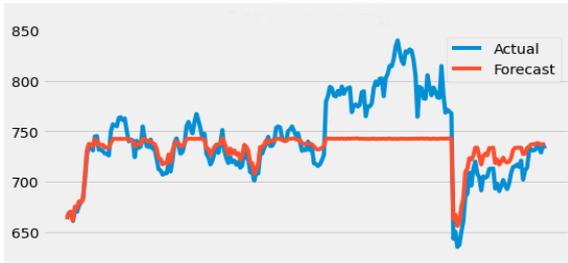

Fig. 3 (d) Actual and Forecasted Stock Prices of Infosys using Random Forest

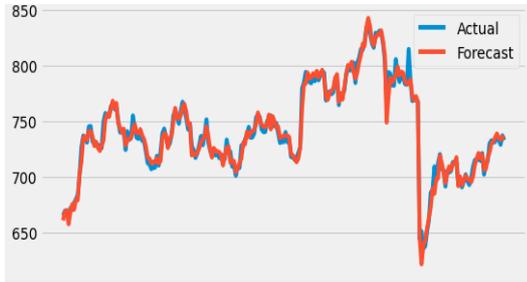

Fig. 3 (e) Actual and Forecasted Stock Prices of Infosys using MARS

Table V. Results of Actual and Forecasted Stock Prices of Infosys using ML-based models

| Models | Training Data | | Testing Data | |
|---|---|---|---|---|
| Random Forest | RMSE/ mean of *close* | 0.003 | RMSE/ mean of *close* | 0.041 |
| MARS | RMSE/ mean of *close* | 0.011 | RMSE/ mean of *close* | 0.008 |

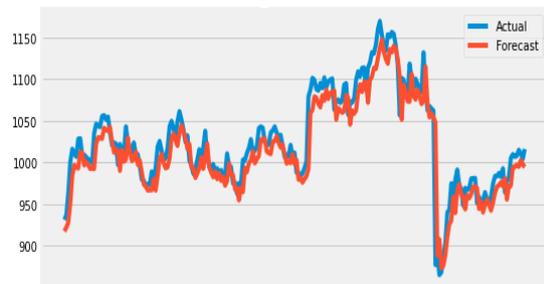

Fig. 3 (e) Actual and Forecasted Stock Prices of Infosys using Simple RNN

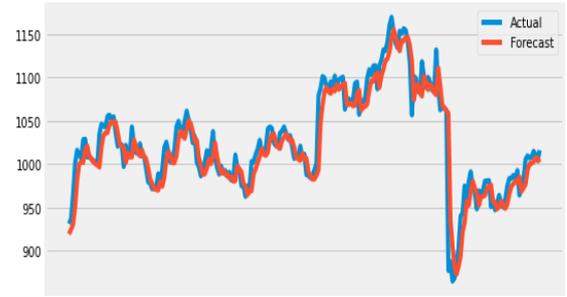

Fig. 3 (g) Actual and Forecasted Stock Prices of Infosys using LSTM

Table VII Results of Actual and Forecasted Stock Prices of Infosys using DL based models

| Models | Training Data | | Testing Data | |
|---|---|---|---|---|
| RNN | RMSE/ mean of *close* | 0.0235 | RMSE/ mean of *close* | 0.022 |
| LSTM | RMSE/ mean of *close* | 0.0244 | RMSE/ mean of *close* | 0.021 |

## B. Banking Sector

In the Banking sector, the *close* value of the stock price of ICICI was considered in our study. Six models were applied to get the forecasted stock price. The output of all the models is shown along with the table of each model, comprising the evaluation metric.

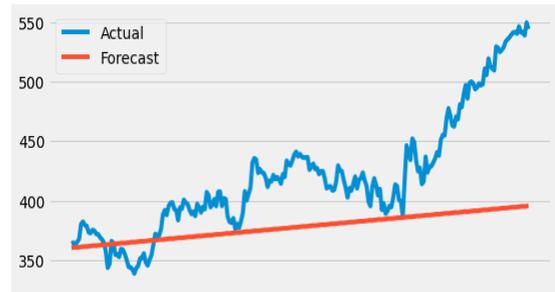

Fig. 4 (a) Actual and Forecasted Stock Prices of ICICI using Holt-Winters Exponential Smoothing

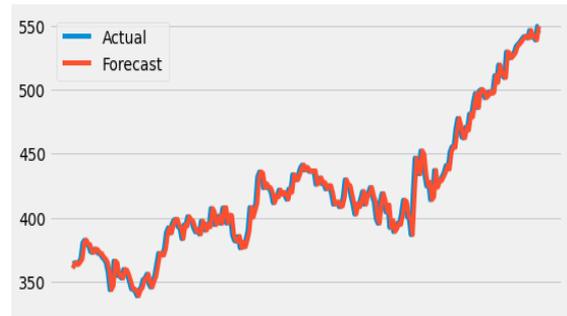

Fig. 4 (b) Actual and Forecasted Stock Prices of ICICI using ARIMA

Table X. Results of Actual and Forecasted Stock Prices of ICICI using TS and econometric based models

| Models | Training Data | | Testing Data | |
|---|---|---|---|---|
| Holt-Winters | RMSE/ mean of *close* | 0.1611 | RMSE/ mean of *close* | 0.142 |
| ARIMA | RMSE/ mean of *close* | 0.036 | RMSE/ mean of *close* | 0.018 |

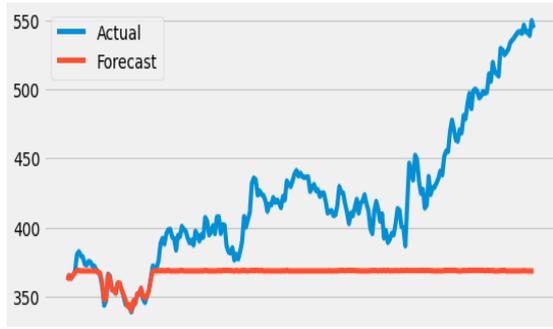

Fig. 4 (d) Actual and Forecasted Stock Prices of ICICI using Random Forest

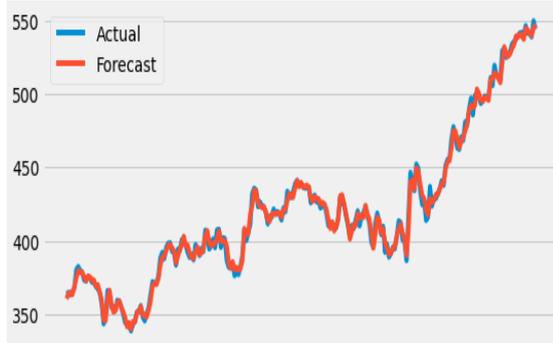

Fig. 4 (e) Actual and Forecasted Stock Prices of ICICI using MARS

Table XII. Results of Actual and Forecasted Stock Prices of ICICI using ML-based models

| Models | Training Data | | Testing Data | |
|---|---|---|---|---|
| Random Forest | RMSE/ mean of *close* | 0.003 | RMSE/ mean of *close* | 0.171 |
| MARS | RMSE/ mean of *close* | 0.011 | RMSE/ mean of *close* | 0.007 |

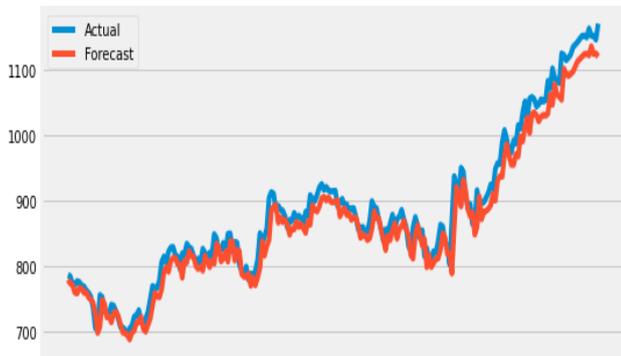

Fig. 4 (f) S Actual and Forecasted Stock Prices of ICICI using Simple-RNN

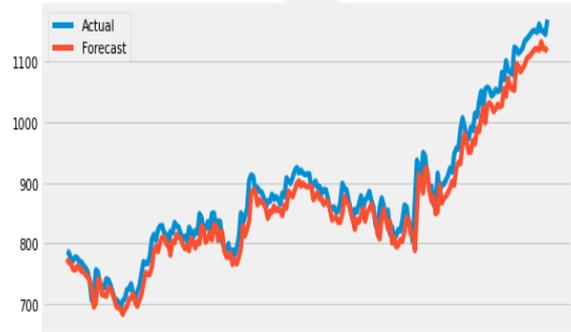

Fig. 4 (g) Actual and Forecasted Stock Prices of ICICI using LSTM

Table XIV. Results of Actual and Forecasted Stock Prices of ICICI using DL based Models

| Models | Training Data | | Testing Data | |
|---|---|---|---|---|
| RNN | RMSE/ mean of *close* | 0.03 | RMSE/ mean of *close* | 0.028 |
| LSTM | RMSE/ mean of *close* | 0.033 | RMSE/ mean of *close* | 0.031 |

*C. Health Sector*

In the Health sector, the close value of the stock price of SUN PHARMA was considered in our study. Six models were applied to get the forecasted stock price. The output of all the models is shown along with the table of each model, comprising the evaluation metric.

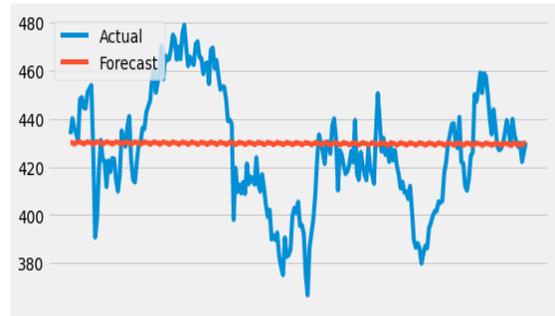

Fig. 5 (a) Actual and Forecasted Stock Prices of SUN PHARMA using Holt-Winters Exponential Smoothing

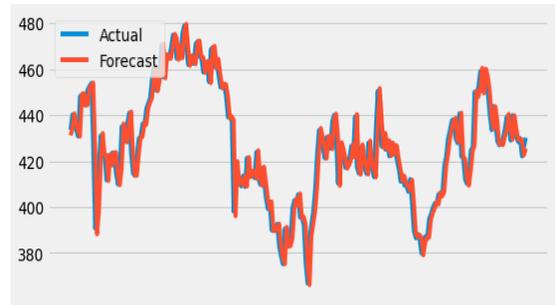

Fig. 5 (b) Actual and Forecasted Stock Prices of SUN PHARMA using ARIMA

Table XVII. Results of Actual and Forecasted Stock Prices of SUN PHARMA using TS and econometric based models

| Models | Training Data | | Testing Data | |
|---|---|---|---|---|
| Holt-Winters | RMSE/ mean of *close* | 0.807 | RMSE/ mean of *close* | 0.855 |
| ARIMA | RMSE/ mean of *close* | 0.558 | RMSE/ mean of *close* | 0.020 |

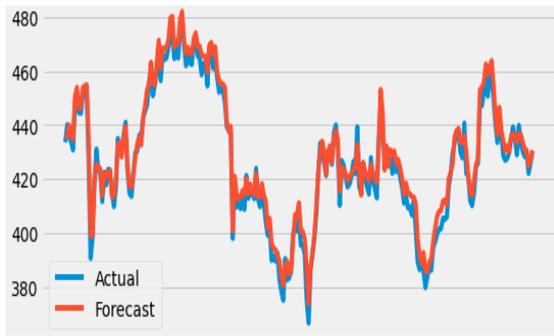

Fig. 5 (d) Actual and Forecasted Stock Prices of SUN PHARMA using Random Forest

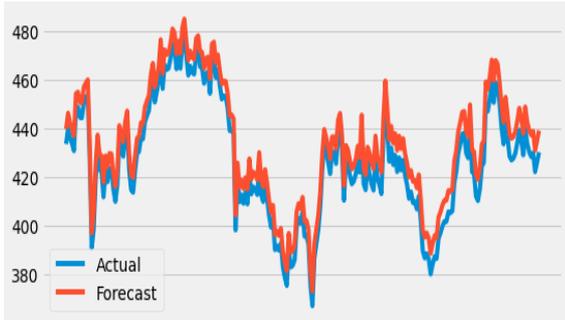

Fig. 5 (e) Actual and Forecasted Stock Prices of SUN PHARMA using MARS

Table XIX. Results of Actual and Forecasted Stock Prices of SUN PHARMA using ML-based models

| Models | Training Data | | Testing Data | |
|---|---|---|---|---|
| Random Forest | RMSE/ mean of *close* | 0.002 | RMSE/ mean of *close* | 0.009 |
| MARS | RMSE/ mean of *close* | 0.007 | RMSE/ mean of *close* | 0.017 |

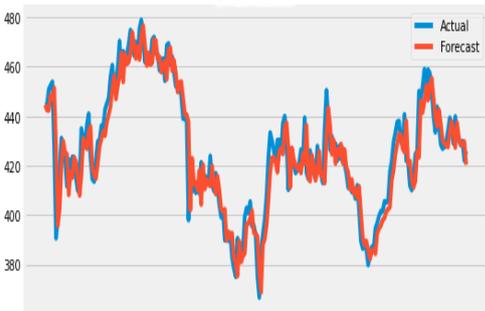

Fig. 5 (f) Actual and Forecasted Stock Prices of SUN PHARMA using Simple-RNN

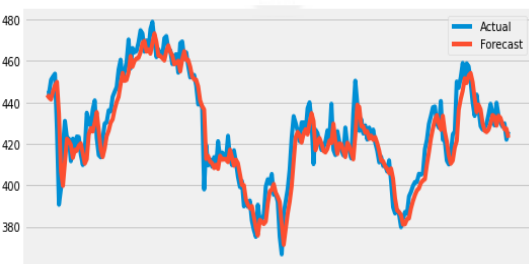

Fig. 5 (f) Actual and Forecasted Stock Prices of SUN PHARMA using LSTM

Table XXI. Results of Actual and Forecasted Stock Prices of SUNPHARMA using DL based Models

| Models | Training Data | | Testing Data | |
|---|---|---|---|---|
| RNN | RMSE/ mean of *close* | 0.027 | RMSE/ mean of *close* | 0.021 |
| LSTM | RMSE/ mean of *close* | 0.031 | RMSE/ mean of *close* | 0.022 |

### D. Comparative study of the performance of all used models in all three sectors

A comparative study based on the performance of each and every model in all three sectors has been studied by evaluating the ratio of RMSE and mean of the *close* value of stock price and has been framed in a tabular format.

Table XXII. Performance Results of Actual and Forecasted Stock Prices of three sectors using all models

| Stocks | Models | RMSE/ mean of the *close* of the test set |
|---|---|---|
| IT Sector | Holt-Winters | 0.089 |
| | ARIMA | 0.018 |
| | Random Forest | 0.041 |
| | MARS | 0.0079 |
| | RNN | 0.0224 |
| | LSTM | 0.021 |
| Banking Sector | Holt-Winters | 0.142 |
| | ARIMA | 0.018 |
| | Random Forest | 0.1714 |
| | MARS | 0.0072 |
| | RNN | 0.0274 |
| | LSTM | 0.0311 |
| Health Sector | Holt-Winters | 0.056 |
| | ARIMA | 0.020 |
| | Random Forest | 0.009 |
| | MARS | 0.017 |
| | RNN | 0.0209 |
| | LSTM | 0.022 |

It can be clearly observed from the above table that Holt-Winters Exponential Smoothing technique performed best in the case of the Health Sector (on SUN PHARMA data) and performed unsatisfactorily in the case of the Banking Sector (on ICICI data) as trend, level, and seasonality gets involved in the data of Health Sector. ARIMA performed well in the case of the IT sector (on ICICI data) and Banking sector (ICICI) and performed poorly for Health Sector (on SUN PHARMA data) as the data of the IT sector was stationary. Random Forest performed best in the case of the Health Sector (on the SUN PHARMA data) and performed poorly in the case of the Banking Sector (on the ICICI data) because the Health Sector data was highly dimensional and imbalanced data was easily handled by Random Forest. Random Forest helped in selecting important features also while sales forecasting. MARS model performed best in the case of the Banking Sector (on ICICI data) and performed poorly in the case of the Health Sector (on SUN PHARMA data) due to its ability to adapt nonlinearity and build a robust nonlinear model, which helped in tracking nonlinearity of Banking data. RNN model performed best in the case of the Health Sector (on the SUN PHARMA data) and performed poorly in the case of the Banking Sector (on the ICICI data) as Health Sector data was sequential. LSTM model performed best in the case of the IT Sector (on the Infosys data) and performed poorly in the case of the Banking Sector (on the ICICI data) as the IT sector was sequential. LSTM performed better over RNN, as it has a long-term memory and it handles vanishing gradient problems well.

VI. CONCLUSION

Several approaches were taken in this paper for forecasting stock price values. One time Series model, one econometric model, two ML models, and two DL-based models were considered in this work to forecast the stock price of three different sectors. The models were devised, trained, and optimized using the training dataset for the period of January 2004 to December 2018 daily. The trained model was run on the test dataset of the year 2019. While the high level of accuracy was obtained from all the applied models, ARIMA was proved to be the best among time series and econometric models, MARS was proved to be the most accurate between machine learning models, and LSTM was proved to be the best in deep learning models. The dexterity of forecasting sequential data made ARIMA producing accurate results. The feature selection ability and adopting nonlinearity from a dataset of MARS made it perform extremely well. On the other hand, the ability to design a model to deal with complex sequential data due to not suffering from *vanishing gradient* and *exploding gradient* problems made LSTM produce an excellent result. We plan to add CNN-based stock price forecasting in our future work because of its ability to extract a large number of features and then subsampling them, and also its high execution speed will result in high predicting accuracy in forecasting.